\documentclass[final,12pt,authoryear]{elsarticle}
\usepackage{graphicx}
\usepackage{epstopdf}
\usepackage[utf8]{inputenc}

\usepackage[margin=1in]{geometry}

\makeatletter
\def\ps@pprintTitle{%
 \let\@oddhead\@empty
 \let\@evenhead\@empty
 \def\@oddfoot{}%
 \let\@evenfoot\@oddfoot}
\makeatother

\usepackage{setspace}
\usepackage{natbib}

\usepackage{bm}
\usepackage{mathrsfs}  
\usepackage{amssymb}
\usepackage{amsmath}
\usepackage{amsthm}
\usepackage{ifpdf}
\usepackage{longtable}
\usepackage[font=small,labelfont=bf]{caption}

\newtheorem{assumption}{Assumption}
\newtheorem{theorem}{Theorem}

\usepackage{booktabs} 

\usepackage{threeparttable}

\usepackage{appendix}

\usepackage{float}
\usepackage{xcolor}

\usepackage{hyperref}
\usepackage{multirow} 
\usepackage{adjustbox}

\begin{document}
\begin{frontmatter}

\title{\LARGE{Threshold Regression in Heterogeneous Panel Data
with Interactive Fixed Effects}\tnoteref{ack}}
\tnotetext[ack]{Acknowledgements: The authors would like to thank Otilia Boldea, Chihwa Kao, Oliver Linton, Esfandiar Maasoumi, Hashem Pesaran, Liangjun Su, Robert Taylor, Lorenzo Trapani, Elias Tzavalis, Joakim Westerlund, and seminar participants at the University of Birmingham, the University of Leicester, the University of York, the Athens University of Economics and Business, the 2023 Italian Congress of Econometrics and Empirical Economics, and the 28th International Panel Data Conference for their valuable comments and suggestions.}

\author{$\text{Marco Barassi}^{a}$}
\author{$\text{Yiannis Karavias}^{b,\ast} $}
\author{$\text{Chongxian Zhu}^{a}$}
\hspace{.2cm}\\
\address{ $\text{University of Birmingham}^{a}$\\
     $\text{Brunel University of London}^{b}$}

\cortext[cor1]{Corresponding Author. Department of Economics, Finance and Accounting, Brunel University of London, Uxbridge, UB8 3PH, London, UK. E-mail address: \texttt{yiannis.karavias@brunel.ac.uk}.}

\begin{abstract}
This paper introduces unit-specific heterogeneity in panel data threshold regression. We develop the asymptotic theory for models with heterogeneous thresholds, heterogeneous slope coefficients, and  interactive fixed effects. The estimation methodology employs the Common Correlated Effects approach, which is able to handle heterogeneous parameters while maintaining computational simplicity. We also propose a semi-homogeneous model with heterogeneous slopes but a common threshold, revealing novel mean group estimator convergence rates due to the interaction of heterogeneity with the shrinking threshold assumption. Tests for linearity are provided, as well as a modified information criterion which can select between the fully heterogeneous and semi-homogeneous models. Monte Carlo simulations demonstrate the good performance of the new methods in small samples. The new theory is used to examine the Feldstein-Horioka puzzle, showing that threshold nonlinearity with respect to trade openness occurs only in a small subset of countries.\\
\end{abstract}


\end{frontmatter}
\pagebreak 

\newcounter{subsubsubsection}[subsubsection]
\renewcommand\thesubsubsubsection{\thesubsubsection.\arabic{subsubsubsection}}
\renewcommand\theparagraph{\thesubsubsubsection.\arabic{paragraph}}

\doublespacing

\section{Introduction}
Threshold regression is one of the most prominent classes of non-linear models used in econometrics. Its principal advantage lies in its intuitive modelling of the regime-switching mechanism, allowing model parameters to change when an observed variable crosses a certain value. As a result, threshold regression can be viewed as a subclass of time-varying parameter models in which the transition mechanism between regimes is explicitly specified rather than left implicit. This framework can be used to characterize regime switching between low- and high-inflation environments, expansions and recessions, or more broadly, ``good times'' and ``bad times''. In the contemporary economic environment, characterized by major shocks such as the COVID-19 pandemic, the war in Ukraine, the United Kingdom’s withdrawal from the European Union, and the recent intensification of the geopolitical and economic competition, such regime shifts are particularly relevant, as changes in the underlying economic dynamics may have pushed key variables over certain thresholds and models into alternative regimes. Threshold regression was introduced in panel data by \citet{hansen1999threshold} and has since been an active field of theoretical and empirical research because  the pooled information across units helps with the identification of the different regimes and the efficient estimation of the threshold parameter. 

A key challenge which arises when many cross-sectional units are pooled together is that of heterogeneity. Heterogeneity arises naturally and the greater the number of units, the more likely the model parameters, such as intercepts, slopes and thresholds, will vary between units. This is a fact that is well recognized in the panel data literature, leading to specialized estimators; see, e.g., the contributions in \citet{swamy1970efficient}, 
\citet{pesaran2006estimation}, \citet{fernandez2013panel}, 
\citet{gao2020heterogeneous}, \citet{trapani2021inferential}, and \citet{lusu23}, inter alia.   Heterogeneity also arises in unobserved unit characteristics. In microeconomics, wages depend on unobserved soft skills and ability, which are heterogeneous across individuals and are likely to have time-varying prices. In macroeconomics, unobserved common shocks, typically modeled as factors, affect countries in a heterogeneous manner causing varying patterns to economic growth. In finance, unobserved factors load heterogeneously on asset returns. This heterogeneity in the unobserved part of the model is also well recognized in the literature and is modeled by interactive fixed effects (IFE), which, as their name suggests, are the inner product of a vector of individual effects with a vector of common time effects. IFE is a generalization of the standard two-way fixed effects and a more flexible and empirically relevant way of capturing unobserved heterogeneity, see e.g. \cite{DitzenKaravias2025} for a discussion and a review of the literature.

Motivated by the above, this paper aims to address the issue of heterogeneity in panel data threshold regression. Specifically, we provide a comprehensive asymptotic theory for estimation and testing in a panel threshold regression model with two distinct features: i) heterogeneous threshold and slope parameters across units and ii) interactive fixed effects. This is the first paper which allows for this type of analysis in threshold regression.

First, we consider a model with fully heterogeneous slope coefficients and thresholds. Although such a model may appear to be estimable on a series-by-series basis by treating each unit as a univariate time series and by applying standard threshold regression methods \citep{hansen2000sample}, this approach is invalid in our setting due to the factor structure in the errors induced by IFE. While series-by-series estimation could, in principle, be conducted under the framework of \cite{andrews2005cross} which allows for common shocks in errors, this framework imposes restrictive assumptions-most notably, independence of observations in time and independence between factors in the errors and the regressors-and does not address nonlinear models such as threshold regressions, making extensions nontrivial and potentially intractable. Instead, we exploit the large cross-sectional dimension of the panel: by using cross-sectional averages \citep{pesaran2006estimation}, we asymptotically remove the IFE and thereby enable consistent series-by-series estimation of the heterogeneous slopes and thresholds. Afterwards, cross-sectional aggregates can be used to summarize the estimated parameters, when $N$ is large enough so that the the individual-specific parameters become less interesting. Examples include the means and medians of the heterogeneous estimates, distributional features such as quantiles and percentiles, group averages for known groups of interest, or shares of units with given sign or class of coefficient magnitude. 

We also consider a second model; one in which slope coefficients are heterogeneous across units, while the threshold parameter is now common to all units. We refer to this as the semi-homogeneous model. This model has been briefly considered in the literature \citep{chudik2017there} but here we show that it can lack identification if the threshold variable does not have a common support across units. Then, we demonstrate that it offers efficiency gains and faster, non-standard, rates of convergence for both the pooled threshold estimator and for the mean-group slope coefficient estimator. The latter is a novel finding, not documented elsewhere, and it appears because the, standard in the literature, ``shrinking threshold'' assumption (see, e.g., \cite{hansen2000sample}) interacts with slope heterogeneity in a way which leads to faster rates of convergence. Finally, a novel modified information criterion which allows distinguishing between the fully heterogeneous and the semi-homogeneous models is provided.

The heterogeneity of the threshold and slope parameters, together with IFE, all have important implications for model estimation requiring appropriate estimators; with potential candidates including generalized method of moments estimators \citep{ahn2013panel}, the Common Correlated Effects (CCE) estimator \citep{pesaran2006estimation}, the Principal Components estimator \citep{bai2009panel}, the two-stage instrumental variables estimator \citep{norkute2021instrumental}, the Post-Nuclear Norm Regularized estimator of \citet{moon2018nuclear}, and Lasso-type shrinkage methods \citep{su2016identifying}. While the above estimators are designed for IFE, not all of them can deal with heterogeneous coefficients. The CCE estimator stands out because it is general enough to allow for heterogeneous coefficients, is analytically tractable, has excellent small sample properties, and is computationally fast. The latter property is important in threshold regression with large panels, where recursive regressions and the bootstrap are necessary, thus making CCE the estimation method of choice. Furthermore, the CCE estimator has been extensively studied in terms of its assumptions, for example in terms of its rank condition or the presence of additional or distinct factors \citep{juodis2022,DeVosStauskas2024}. \cite{juodisreese2026} summarize advice on the use of CCE. Yet despite these properties, the application of CCE in threshold regression is not straightforward because the dependent variable cross-section averages include ``threshold factors'' in the errors, increasing the dimension of the factor space to be estimated. Hence, we show that the vanila CCE estimator can be applied only in the case where the number of factors is exactly equal to $1$. Because this is an important restriction, we employ a modified CCE which does not include cross-section averages from the dependent variable.

In terms of threshold regression in heterogeneous panel data models, the paper closest to us is that of \citet{chudik2017there}, which considers heterogeneous slopes and IFE, but not heterogeneous thresholds. Furthermore, it is mostly focused on the specific empirical application and does not provide any asymptotic theory supporting the estimation methodology. The present paper fills these gaps. \citet{miao2020panel2} assumes that the units belong to a small number of groups and that both the slope and threshold parameters vary between the groups. However, their analysis applies only to models with fixed effects and not to the full IFE model considered here. Furthermore, the empirical application is focused on estimating the number of underlying groups and group membership, which is different from the type of parameter heterogeneity considered here. Therefore, the present contribution is clearly distinct. \citet{miao2020panel1} consider panel threshold regression with IFE but restrict the parameters to be homogeneous across units. \citet{haciouglu2021common} consider a smooth transition model with heterogeneous coefficients and IFE. Other contributions in the area include panel kink regression with covariate-dependent threshold \citep{yangzhang20} and threshold regression in dynamic panel data models with a short time dimension \citep{seo2016dynamic}. However, these all assume homogeneous coefficients and simple fixed effects.

We conclude by applying the new methodology to examine one of the most important macroeconomic problems, the Feldstein-Horioka puzzle \citep{feldstein1980domestic}. Our new model allows for heterogeneous coefficients and thresholds and for cross-sectional dependence, which are data features well documented in country-level data \citep{chudik2017there}. Previous research has considered nonlinear effects with respect to trade openness \citep{feldstein1980domestic} but with mixed results. We confirm the existence of the puzzle, while threshold effects in the high-trade-openness regime are found only for a small subset of countries. It seems that the threshold effects found elsewhere in the literature could be driven by only a few countries in the sample as opposed to being a general phenomenon. 

The remainder of the paper is organized as follows. Section 2 introduces the fully heterogeneous model with heterogeneous slopes and thresholds. Section 3 develops the estimation strategy. Sections 4 and 5 provide the assumptions and asymptotic theory, respectively. Section 6 introduces the semi-homogeneous model in which the thresholds are common across units. Section 7 discusses diagnostics, while Section 8 applies the new methodology to study the Feldstein-Horioka puzzle. Section 9 concludes. The supplementary online appendix contains the bootstrap algorithms for the linearity tests, the information criterion for selecting between models, extensive Monte Carlo simulations, additional results for the empirical application, and all mathematical proofs.
\setcounter{assumption}{0}
\renewcommand{\theassumption}{A.\arabic{assumption}}

\section{A Fully Heterogeneous Threshold Model}\label{section:model}
Consider the following model with $N$ units and $T$ time-series observations. The response variable of the $i^{\text{th}}$ unit, observed at time $t$, $y_{i,t}$ is given by: 
\begin{equation} \label{basicmodel}
	y_{i,t} = \beta_{i}'x_{i,t} + \delta_{i}'w_{i,t} \mathbb{I}\{q_{i,t} \leq \gamma_{i} \}+ e_{i,t}, \quad \text{$i = 1,...,N$, $t = 1,...,T$},
\end{equation} 
where $x_{i,t}$ is $K \times 1$ vector of observable regressors, and $\beta_{i}$ is a $K \times 1$ vector of heterogeneous slope coefficients, which can be different for each unit $i$. Additionally, let $w_{i,t} = R'x_{i,t}$ be an $r \times 1$ subset of the regressors in $x_{i,t}$. The matrix $R$ is a $K \times r$ selection matrix of zeros and ones, with full column rank $r$, that selects elements of $x_{i,t}$ whose coefficients are subject to the threshold effect. $R$ is known to the researcher. If $R = I_K$, where $I_K$ is the $K-$th order identity matrix, then all $K$ regressors in $x_{i,t}$ have the threshold effect and the model is called a pure threshold model, while if $R = (I_r,0_{r \times (K-r)})'$, then only the first $r$ regressors in $x_{i,t}$ are affected by the threshold, and the model is called a partial threshold model. An additional classification is introduced here: the model is called fully heterogeneous when $\gamma_i$ varies across units and semi-homogeneous when $\gamma_i=\gamma$ for all $i$. To the best of our knowledge, this is the first paper which allows $\gamma_i$ to vary across units. This section proceeds with the fully heterogeneous model.

A key characteristic of \eqref{basicmodel} is that the effect of the regressors on the dependent variable is allowed to vary between two regimes which are identified by the indicator function $\mathbb{I}$, which takes the value 1 when $\{q_{i,t} \leq \gamma_i \}$ and 0 otherwise. The variable $q_{i,t}$ is a scalar that may belong to $x_{i,t}$, and $\gamma_i$ is the threshold parameter that defines the two different regimes for unit $i$. When $q_{i,t} > \gamma_i$ the effect of $x_{i,t}$ on $y_{i,t}$ is $\beta_i$. This is often called the ``high regime''. There is also a ``low regime'' with $q_{i,t} \leq \gamma_i$, where the coefficient of the variables is $\beta_i + \delta_i$. The model is suitable to identify the different equilibria which can arise when the response of the dependent variable is different across periods. Examples of the ``low regime'' include periods of economic and financial turmoil and distress, such as poor stock market performance or economic crises, or periods of unfavourable economic outlook and low sentiment compared to normal times, namely the ``high regime''. Similarly, it is possible to employ $\mathbb{I}\{q_{i,t} > \gamma_i \}$ in \eqref{basicmodel}, for example, to capture regimes where inflation or interest rates are above a certain threshold in monetary policy models (high inflation regimes). The indicator function definition only affects the interpretation of the $\delta_i$ coefficients.   

The model in \eqref{basicmodel} is sufficiently general to render existing panel models as special cases; $(i)$ the simple linear heterogeneous panel model of \citet{swamy1970efficient} corresponds to the case where $\delta_{i} = 0$ for all $is$ and $(ii)$ the homogeneous threshold panel models of \citet{hansen1999threshold} and \citet{miao2020panel1} correspond to the case where $\beta_{i} = \beta$, $\delta_{i} = \delta$ and $\gamma_i=\gamma$.

The errors  $e_{i,t}$ contain the unobserved heterogeneity, which has the general form of IFE or a multi-factor error structure as it is sometimes called in the literature \citep{pesaran2006estimation}:
\begin{equation}
	    e_{i,t} = \lambda_{i}'f_t + \varepsilon_{i,t}, \label{or_eq7}
\end{equation}
where $f_t$ is a $m \times 1$ vector containing $m$ unobserved factors which are common to all units, $\lambda_{i}$ are the heterogeneous common factor loadings, and $\varepsilon_{i,t}$ are the remaining idiosyncratic errors. The unobserved heterogeneity formulation in \eqref{or_eq7}, encompasses all standard panel data models. If $f_t =1$ then we have the standard fixed effect (FE) model, while for $f_t=(1,\tau_t)'$ and $\lambda_i = (\alpha_i, 1)'$ we have the two-way fixed effect (TWFE). The unobserved common factors $f_t$ can capture the price of an unobserved skill which changes in time or the price of an input for production, or even aggregate shocks to a particular market or economy. Because all units are affected by these factors, the factor structure is the source of cross-sectional dependence between units. 
 
The IFE can be correlated to the regressors. We follow \citet{pesaran2006estimation} and assume that this relationship is linear:  
\begin{equation}
	    x_{i,t} = \Pi_i'f_t + \xi_{i,t}, \label{or_eq8}
\end{equation}
where $\Pi_i$ is $m \times K$ fixed factor loading matrix and $\xi_{i,t}$ is a $K \times 1$ is the idiosyncratic part. In the following we will assume the empirically relevant scenario that $q_{i,t}$ is a variable from $x_{i,t}$ and hence it is also described by the above process. The model \eqref{or_eq8} implies that $y_{i,t}$ and $x_{i,t}$ load on the same factors $f_t$. This assumption is not strictly required, as distinct factors can be handled by using the bootstrap of \citep{DeVosStauskas2024}.

\setcounter{remark}{0}
\section{Fully Heterogeneous Model Estimation}\label{section:estimation}
The existence of factors $f_t$ in $x_{i,t}$ and $e_{i,t}$ causes endogeneity and makes the fixed-effect estimator inconsistent. Instead, we employ the CCE estimator, which we later show is consistent. To present the estimator, we stack the models in \eqref{basicmodel}, \eqref{or_eq7}, and \eqref{or_eq8} across the time dimension. Letting $w_{i,t} (\gamma_i) = w_{i,t} \mathbb{I}\{q_{i,t} \leq \gamma_i \} $, the stacked models become:
\begin{align} \label{or_eq9}
 y_i&= X_i \beta_i + W_i(\gamma_i)\delta_i+ e_i,\\
 \label{or_eq11} X_i &= F \Pi_i + \xi_i,\\
 \label{or_eq10}e_i &= F \lambda_i + \varepsilon_i,
\end{align}
where, $y_i= (y_{i,1}, y_{i,2}, \ldots,y_{i,T})'$ is a $T \times 1$ vector,  $X_i= (x_{i,1}, x_{i,2}, \ldots,x_{i,T})'$, and also $\xi_i = (\xi_{i,1}, \xi_{i,2}, \ldots,\xi_{i,T})'$ are $T \times K$ matrices, while $W_i (\gamma_i)= (w_{i,1}(\gamma_i), w_{i,2}(\gamma_i), \ldots,w_{i,T}(\gamma_i))'$ is $T \times r$. The errors $e_i= (e_{i,1}, e_{i,2}, \ldots,e_{i,T})'$ and $\varepsilon_i = (\varepsilon_{i,1}, \varepsilon_{i,2}, \ldots,\varepsilon_{i,T})'$ are $T\times 1$ vectors. Finally, $F = (f_1, f_2, \dots, f_T)'$ is a $T \times m$ matrix.

The model in \eqref{or_eq9}, is linear in the parameters $\beta_i$ and $\delta_i$ and non-linear in the parameter $\gamma_i$. For now, let the $\gamma_i$'s be known. The model is therefore linear and can be estimated using a variant of the CCE estimator in \citet{pesaran2006estimation}, adapted as in \citet{karavias2022structural}. The key idea is that cross-sectional averages of $X_i$ can be used to consistently estimate the space spanned by unknown factors in \eqref{or_eq10}. This is an alternative to using principal components for estimating the factors; see, e.g. \cite{bai2009panel}, \cite{norkute2021instrumental} and \citet{westerlundurbain2015}. The key benefits of CCE is that it does not require estimating the number of factors, which can be a difficult task \citep{moon2018nuclear}, and it offers better small sample performance \citep{westerlundurbain2015}. Denote the cross-sectional average as $\bar A=\sum_{i=1}^N A_i$ for any $A_i$. Also, for any $T$-rowed matrix $A$, define the projection matrix $M_{A}=I_T-A\left(A'A\right)^{-1}A'$. Then, the first step of estimation involves premultiplying \eqref{or_eq9} by $M_{\bar{X}}=I_T-\bar X\left(\bar X'\bar X\right)^{-1}\bar X'$, after which the transformed model becomes:
\begin{equation}\label{or_eq12}
 \tilde{y}_i= \tilde{X}_i \beta_i + \tilde{W}_i(\gamma_i)\delta_i+ \tilde{e}_i,     
\end{equation}
where $\tilde{y}_i = M_{\bar{X}}y_i$, $\tilde{X}_i = M_{\bar{X}}X_i$, $\tilde{W}_i(\gamma_i) = M_{\bar{X}}W_i(\gamma_i)$ and $\tilde{e}_i = M_{\bar{X}}e_i$. We will show later that $M_{\bar{X}} e_i=M_{\bar{X}}F\lambda_i+M_{\bar{X}}\varepsilon_i =M_{\bar{X}}\varepsilon_i+ o_{p}(1)$, asymptotically removing the $m$ common factors. Here we only use $\bar X$ to remove the IFE, which is different from using both $\bar y$ and $\bar X$ as in the original CCE estimator of \citet{pesaran2006estimation}. Intuitively, this is because $y_{i,t}$ is now a function of $(f_t',f_t'\mathbb{I}\{q_{i,t} \leq \gamma_i \})'$ and therefore the single cross-section average $\bar y_{t}$ will need to estimate the $m$ threshold factors $f_t'\mathbb{I}\{q_{i,t} \leq \gamma_i \}$, which is possible only if $m=1$. This restriction can be strong in practice and hence we do not employ $\bar y_t$ in removing the factors. 

The model in \eqref{or_eq12} can be rewritten in a more compact form as:
\begin{equation}\label{or_eq13}
 \tilde{y}_i= \tilde{Z}_i(\gamma_i)\theta_i+ \tilde{e}_i,   
\end{equation}
where $\tilde{Z}_i (\gamma_i) = \left( \tilde{X}_i, \tilde{W}_i(\gamma_i) \right) $ and $\theta_i = (\beta_i', \delta_i')'$. 
Let $\gamma_i$ be known, the CCE estimators are:
\begin{align}
        \hat{\theta}_i (\gamma_i)= \left(\tilde{Z}_i (\gamma_i)' \tilde{Z}_i (\gamma_i) \right)^{-1} \tilde{Z}_i (\gamma_i)' \tilde{y}_i \label{or_eq13}
\end{align} 
and more explicitly, the parameter estimates are $\hat{\beta}_i(\gamma_i) = \left(\tilde{X}_i' M_{\tilde{W}_i (\gamma_i)} \tilde{X}_i\right)^{-1} \tilde{X}_i' M_{\tilde{W}_i (\gamma_i)} \tilde{y}_i$ and $\hat{\delta}_i(\gamma_i) = \left(\tilde{W}_i (\gamma_i)' M_{\tilde{X}_i} \tilde{W}_i (\gamma_i)\right)^{-1} \tilde{W}_i (\gamma_i)' M_{\tilde{X}_i } \tilde{y}_i$.

If the $\gamma_i$'s are unknown, as it is usually the case in practice, we follow \citet{chan1993consistency} and estimate them by minimizing the CCE sum of squared residuals for each unit $i$:
\begin{align} 
  \hat{\gamma_i} &=\underset{\gamma_i \in \Gamma_i}{\text{argmin}} \left[\tilde{y}_i- \tilde{Z}_i (\gamma_i) \hat{\theta}_i(\gamma_i) \right]'\left[\tilde{y}_i - \tilde{Z}_i (\gamma_i) \hat{\theta}_i(\gamma_i)     \right] \label{or_eq20}.  
\end{align}

For each $i$, the above sum of squared residuals is a step function for $\gamma_i$ that only has $O(T)$ distinct values. When $T$ is large, \citet{hansen1999threshold} suggests approximating $\Gamma_i$ using a grid search method to save computational time, searching in $\Gamma_i \cap \{q_{1,t},  1 \leq t \leq T\}$. First, sort the distinct values of the observations in the threshold variable $q_{1,t}$, and then trim the top and bottom $1\%$, $5\%, 10\%$, or any other specific percentiles of $q_{i,t}$. Finally, search for $\hat{\gamma_i}$ over the remaining values of $q_{i,t}$. Once the $\hat{\gamma_i}$ have been obtained, the estimators $\hat \theta_i(\hat\gamma_i)$, can be obtained by substituting $\hat{\gamma_i}$ for $\gamma_i$ in \eqref{or_eq13}. We will henceforth denote $\tilde{\beta}_i = \hat{\beta}_i (\hat{\gamma_i})$, $\tilde{\delta}_i = \hat{\delta}_i (\hat{\gamma_i})$ and $\tilde{\theta}_i = (\tilde{\beta}_i , \tilde{\delta}_i)$.

The estimator variances which are necessary for confidence intervals and hypothesis testing are given by $\tilde{V}_{\theta_i}=\hat{V}_{\theta_i}(\hat\gamma_i)=\hat{\Sigma}_i^{-1}(\hat\gamma_i)\hat{S}_{i}(\hat\gamma_i)\hat{\Sigma}_i^{-1}(\hat\gamma_i)$, where $\hat{\Sigma}_i (\hat\gamma_i)= T^{-1} \Tilde{Z}_i(\hat{\gamma_i})'\Tilde{Z}_i(\hat{\gamma_i})$ and  $\hat{S}_{i}(\hat\gamma_i)  = T^{-1}\Tilde{Z}_i(\hat{\gamma}_i)'diag(\hat{\varepsilon}_i\hat{\varepsilon}_i')\Tilde{Z}_i(\hat{\gamma}_i)$ in the case of independent in time $\varepsilon_{i,t}$, or
    \begin{equation} \label{NW_eq}
        \hat{S}_{i}(\hat\gamma_i) = \hat{\Lambda}_{i,0}(\hat\gamma_i) + \sum_{j = 1}^{b}\left(1-\frac{j}{b+1}\right)\left(\hat{\Lambda}_{i,j}(\hat\gamma_i) + \hat{\Lambda}_{i,j}'(\hat\gamma_i)\right),
    \end{equation}
    in the presence of serial correlation. $\hat{\Lambda}_{i,j}(\hat\gamma_i) = T^{-1}\sum_{t = j + 1}^{T}\hat{\varepsilon}_{i,t}(\hat{\gamma}_i)\hat{\varepsilon}_{i,t - j}(\hat{\gamma}_i)\tilde{z}_{i,t}(\hat{\gamma}_i)\tilde{z}_{i,t-j}(\hat{\gamma}_i)'$, where $\tilde{z}_{i,t-j}'$ is the $(t-j)$-th row of $\tilde Z_i(\hat\gamma_i)$, and $   \hat{\varepsilon}_i(\hat\gamma_i) = \hat{\tilde{e}}_i(\hat\gamma_i)=\Tilde{y}_i - \Tilde{Z}_i(\hat{\gamma}_i)\hat{\theta}_i(\hat{\gamma}_i)$. Finally, the bandwidth can be $b = \lfloor  T^{1/4} \rfloor $ or selected by any other appropriate rule.

\setcounter{remark}{0}

\section{Fully Heterogeneous Model Assumptions}
This section presents the main assumptions under which we develop the asymptotic theory. In the following, we consider a pure threshold model, i.e. $R = I_{K}$, to simplify notation, and in which case, $w_{i,t} = x_{i,t}$, $w_{i,t}(\gamma_i) = x_{i,t}(\gamma_i)$, and $W_i(\gamma_i)=X_i(\gamma_i)$. Define also, $\Tilde{X}_i(\gamma_i^1,\gamma_i^2) = \Tilde{X}_i(\gamma_i^1) - \Tilde{X}_i(\gamma_i^2)$ for any $\gamma_i^1,\gamma_i^2 \in \Gamma_i$ and $d_{i,t}(\gamma_i) = \mathbb{I} \{q_{i,t} \leq \gamma_i \}$. All the results below also hold for the partial threshold model. 

The letter $C$ stands for a universal finite positive constant. $||A||$ denotes the Frobenius norm, while $l_{min}(A)$ denotes the smallest eigenvalue of $A$. $diag(A)$ denotes a diagonal matrix consisting of the main diagonal elements of the matrix $A$. For a square matrix $A$, $A > 0$ means that $A$ is positive definite. The symbol $ \stackrel{p}{\rightarrow}$ denotes convergence in probability, $\hspace{0.1cm} \stackrel{d}{\rightarrow} \hspace{0.1cm} $ convergence in distribution, $\Rightarrow$ weak convergence with respect to the uniform metric, and $plim$ probability limit. $(N,T) \rightarrow \infty$ denotes that both $N$ and $T$ tend to infinity together. Let $\mathscr{D} = \sigma(F)$ be the minimal sigma-field generated from the factor structure $F$. Let $\mathbb{P}_{\mathscr{D}}(A) = \mathbb{P}({A|\mathscr{D}})$ and $E_{\mathscr{D}}(A) = E(A|\mathscr{D})$. We use the superscript $0$ to denote the true parameter values. In particular, the true coefficients are denoted by $\theta^0_i= (\beta^{0'}_i,\delta^{0'}_i)'$ and the true threshold parameters by $\gamma^0_i$ for $i = 1,...,N$.  

\setcounter{assumption}{0}
\renewcommand{\theassumption}{H.\arabic{assumption}}

\begin{assumption}[Common Factors]\label{ass:commonfactors}
i) $f_t$ is covariance stationary with absolute summable autocovariances, distributed independently of $\varepsilon_{i,t'}$ and $\xi_{i,t'}$ for all $i,t,t'$;
ii) $E||f_{t}||^{4+\epsilon} < \infty$ for some $\epsilon > 0$; iii) $\frac{1}{T}\sum_{t=1}^{T}f_tf_t' \stackrel{p}{\rightarrow} \Sigma_f > 0$ for some $m \times m$ matrix $\Sigma_f$ as $T \rightarrow \infty$.
\end{assumption}

\begin{assumption}[Strong Factors and Fixed Factor Loadings]\label{ass:factorloadings}    
    i) $Rank(\bar{\Pi}) = m \leq K$ for all $N$, including $N \rightarrow \infty$; ii) $||\bar{\Pi} || < \infty$, and $sup_i ||\Pi_i || \leq C < \infty$; iii) $||\lambda_{i} || < \infty$ for $i = 1,2,...,N$, and $sup_i ||\lambda_i || \leq C < \infty$.
\end{assumption}

\begin{assumption}[Regressors]\label{ass:regressors} i) $\xi_{i,t}$ are independent across $i$; ii) for each $i$, $\xi_{i,t}$ follows a linear stationary process with absolute summable autocovariances, $\xi_{i,t} = \sum_{l=0}^{\infty}G_{i,l}\psi_{i,t-l}$, where $\psi_{i,t}$ are independent and identically distributed (i.i.d.) random vectors with zero means, identity variance-covariance matrices and finite fourth-order cumulants. In particular, $Var(\xi_{i,t}) = \sum_{l=0}^{\infty} G_{i,l}G_{i,l}' = \Sigma_{\xi,i} \leq \bar{\Sigma}_{\xi} < \infty$.
\end{assumption}

\begin{assumption}[Errors]\label{ass:errors}
i) $\varepsilon_{i,t}$ are independent across $i$;
ii) for each $i$, $\varepsilon_{i,t}$ follows a linear stationary process with absolute summable autocovariances, $\varepsilon_{i,t} = \sum_{l=0}^{\infty}a_{i,l}\varsigma_{i,t-l}$, where $\varsigma_{i,t}$ is i.i.d. random variables with zero means, unit variances and finite fourth-order cumulants. In particular, $Var(\varepsilon_{i,t}) = \sum_{l=0}^{\infty} a_{i,l}^2 = \sigma_{\varepsilon,i}^2 \leq \bar{\sigma}_{\varepsilon}^2 < \infty$.
iii) $\varepsilon_{i,t}$ and $\xi_{j,t'}$ are distributed independently for all $i,j,t,t'$.
\end{assumption}

\begin{assumption}[Continuous Threshold Variable]\label{ass:contthreshvar}
i) $q_{i,t}$ is a continuous threshold variable. Conditionally on $\mathscr{D}$,
$q_{i,t}$ has a conditional probability density function, $f_{i,t,\mathscr{D}}(\gamma)$. ii) $f_{i,t,\mathscr{D}}(\gamma)$ is continuous at $\gamma_i = \gamma_i^0$ and is uniformly bounded:  $sup_{i,t} sup_{\gamma \in \Gamma} f_{i,t,\mathscr{D}}(\gamma) \leq C < \infty$.
iii) For $k>t$, $f_{i,k|t}(\gamma_i^0|\gamma_i^0) < \infty$, where $f_{i,k|t}$ denotes the conditional density of $q_{i,k}$ given $q_{i,t}$. 
\end{assumption}

\begin{assumption}[Identification Condition] \label{ass:identification} Consider the cross-product terms $\mathcal{S}_{1,i}(\gamma_i) = \Tilde{Z}_i(\gamma_i)'\Tilde{Z}_i(\gamma_i)$, $\mathcal{S}_{2,i}(\gamma_i) = \Tilde{Z}_i(\gamma_i)'\Tilde{X}_i(\gamma_i^0,\gamma_i)$, and $\mathcal{S}_{3,i}(\gamma_i) = \Tilde{X}_i(\gamma_i^0,\gamma_i)'\Tilde{X}_i(\gamma_i^0,\gamma_i)$. Then,  
i) $\delta_i \neq 0$ for each unit $i$, ii) $\mathcal{S}_{1,i}(\gamma_i)>0$, and 
iii) there exist constants $\tau_i > 0$, such that, as $T  \rightarrow \infty$: 
\begin{equation}
    \mathbb{P}\left\{\min_{\gamma_i \in \Gamma_i} \lambda_{min}\left[\frac{1}{T}\left(\mathcal{S}_{3,i}(\gamma_i) - \mathcal{S}_{2,i}(\gamma_i)'\mathcal{S}_{1,i}^{-1}(\gamma_i)\mathcal{S}_{2,i}(\gamma_i)\right)\right] \geq \tau_i min \left(1, |\gamma_i - \gamma_i^0|\right) \right\} \stackrel{p}{\rightarrow} 1.
\end{equation}
\end{assumption}

\begin{assumption}[Threshold Parameter Space]\label{ass:threshspace} The threshold parameters $\gamma_i^0\in \Gamma_i=[\underline{\gamma_i},\overline{\gamma_i}]$, where $\Gamma_i$ are compact sets.
\end{assumption}

\begin{assumption}[Full Rank Conditions] \label{ass:fullrankH}
i) Conditionally on $\mathscr{D}$, let $M_{T,i}^{\star} = E_{\mathscr{D}}\left(T^{-1}{X}_i'M_FX_i\right)$, 
$M_{T,i}^{\star}(\gamma_i^1,\gamma_i^2) = E_{\mathscr{D}}\left(T^{-1}X_i(\gamma_i^1)'M_FX_i(\gamma_i^2)\right)$, where for $\gamma_i^1 = \gamma_i^2 = \gamma_i$ let $M_{T,i}^{\star}(\gamma_i)=M_{T,i}^{\star}(\gamma_i^1,\gamma_i^2)$, and $M^{\star\star}_{T,i}(\gamma_i) = E_{\mathscr{D}}\left(T^{-1}X'_iM_FX_i(\gamma_i)\right)$. Then, for all $\gamma_i\in\Gamma_i$:
\begin{equation}
    Q_{1i}(\gamma_i) = \begin{bmatrix}
		M_{T,i}^{\star} & M_{T,i}^{\star\star}(\gamma_i)\\
		M_{T,i}^{\star\star}(\gamma_i) & M_{T,i}^{\star}(\gamma_i)\\
		\end{bmatrix} > 0.
\end{equation}
\end{assumption}

\begin{assumption}[Uniformly Bounded Moments] \label{ass:uniboundmom}
Conditional on $\mathscr{D}$,
    i) $sup_i (M_{T,i}^{\star})^2 \leq C < \infty$; furthermore, ii) $sup_{\gamma_i} sup_{i}\left[M^{\star\star}_{T,i}(\gamma_i)\right]^{2} \leq C < \infty$, $sup_{\gamma_i} sup_i \left[M_{T,i}^{\star}(\gamma_i)\right]^2 \leq C < \infty$; iii) $sup_{\gamma_i} sup_i E(T^{-1}X_i(\gamma_i)'X_i(\gamma_i))^2 \leq C < \infty$.
\end{assumption}

\begin{assumption}[Shrinking Thresholds] \label{ass:shrink_thres}
		For each unit $i$, let $C_{0,i}$ be fixed constants with $C_{0,i}\in \mathbb{R}^{r}$ and $sup_{i}$ $||C_{0,i}|| < \infty$. For $0 < \alpha_i < 1/2$, the threshold effects $\delta^0_{i}$ satisfy $\delta^0_{i} = C_{0,i}T^{-\alpha_i}$ for $i = 1,2,...,N$.
\end{assumption}

\begin{assumption}[Higher Order Moments]\label{ass:hiordmom} Let 
$M_{\mathscr{D},i}(\gamma_i) = T^{-1}$ $\sum_{t=1}^{T}E_{\mathscr{D}}(C_{0,i}'x_{i,t}x_{i,t}'C_{0,i}|q_{i,t} = \gamma_i)f_{i,t,\mathscr{D}}(\gamma_i)$. Then, 
i) there exist $\mathscr{D}$-dependent variables $C_{i,t}^{\mathscr{D}}$ such that
    $sup_{\gamma_i \in \Gamma_i} E_{\mathscr{D}}(||x_{i,t} ||^4 |q_{i,t} = \gamma_i) \leq C_{i,t}^{\mathscr{D}}$,  
     $sup_{\gamma_i \in \Gamma_i} E_{\mathscr{D}}(||x_{i,t}e_{i,t} ||^4 |q_{i,t} = \gamma_i) \leq C_{i,t}^{\mathscr{D}}$, and $\mathbb{P}\Big(T^{-1}\sum_{t=1}^{T}{C_{i,t}^{\mathscr{D}}}^2 \leq C^2 \Big) = 1$, for some $C < \infty$, as $T \rightarrow \infty$.     ii) $M_{\mathscr{D},i}(\gamma_i^0) > 0$ a.s. and continuous at $\gamma_i = \gamma_i^0$. 
    iii) For all $\epsilon_i > 0$, there exist constants $B_i > 0$ and $C_{1i} > 0$, for large enough $T$, we have $ \mathbb{P}\Big(\mathop{inf}_{|\gamma_i - \gamma_i^0| < B_i} M_{\mathscr{D},i}(\gamma_i) > C_{1i} \Big) > 1 - \epsilon_i.$
\end{assumption}
   
\begin{assumption} [Well-behaved Limits]\label{ass:wellbehlim} Let $f_{i,t}(\gamma_i)$ be the unconditional probability density function of $q_{i,t}$, $D_{i,T}(\gamma_i) = T^{-1}\sum_{t=1}^{T}E\left[(C_{0,i}'x_{i,t})^2|q_{i,t} = \gamma_i \right]f_{i,t}(\gamma_i)$, and finally $V_{i,T}(\gamma_i) = T^{-1}\sum_{t=1}^{T}E\left[(C_{0,i}'x_{i,t})^2\varepsilon_{i,t}^2 |q_{i,t} = \gamma_i \right]f_{i,t}(\gamma_i)$. Consider also the limits as $T\to\infty$: $D_i(\gamma_i) = plim_{(N,T) \rightarrow \infty}D_{i,T}(\gamma_i)$ and $V_i(\gamma_i) = plim_{(N,T) \rightarrow \infty}V_{i,T}(\gamma_i)$. Then: i)  $D_i(\gamma_i)$ and $V_i(\gamma_i)$ exist, with $D_i(\gamma^0_i) > 0$, and $V_i(\gamma^0_i) > 0$, and are both continuous at $\gamma_i = \gamma^0_i$. ii) The following variances satisfy:    $Var\left[T^{-1/2}\sum_{t=1}^{T}E_{\mathscr{D}}(||g^{(1)}_{i,t}(\gamma^1_i,\gamma^2_i) ||^2)  \right]  \leq C^\star_i sup_{i,t} E(||g^{(1)}_{i,t}(\gamma^1_i,\gamma^2_i) ||^4),$ and $Var\left[T^{-1/2}\sum_{t=1}^{T}E_{\mathscr{D}}(||g^{(2)}_{i,t}(\gamma^1_i,\gamma^2_i) ||^2) \right] \leq C^\star_i sup_{i,t} E(||g^{(2)}_{i,t}(\gamma^1_i,\gamma^2_i) ||^4),$ for constants $C^{\star}_i$, where $g^{(1)}_{i,t}(\gamma^1_i,\gamma^2_i) = x_{i,t}|d_{i,t}(\gamma^1_i) - d_{i,t}(\gamma^2_i)|$, $ g^{(2)}_{i,t}(\gamma^1_i,\gamma^2_i) = x_{i,t}\varepsilon_{i,t}|d_{i,t}(\gamma^1_i) - d_{i,t}(\gamma^2_i)|$. 
\end{assumption}

Assumption \ref{ass:commonfactors} is standard in the literature, and it is similar to Assumption 1 in \citet{pesaran2006estimation} which excludes nonstationary factors and trends. Part i) of assumption \ref{ass:factorloadings} is the so-called rank condition $rank(\bar\Pi)=m\leq K$, and states that the number of factors must be smaller or at most equal to the number of regressors. It also implies that the factors must be strong. When it comes to the factor loadings, we follow \citet{westerlund2022cce} and assume in parts ii) and iii) that $\lambda_{i}$ and $\Pi_{i}$ are fixed in our setting, unlike \citet{pesaran2006estimation}. By assuming them as constants we avoid imposing further assumptions such as being i.i.d. and independent to other random elements in the model. 

The requirement $m\leq K$ is standard in the CCE literature and appears in all CCE-based methods. Applied research typically finds a small number of factors $m$ in the error term \citep{DitzenKaravias2025}. To further relieve the strain of the rank condition on the number of regressors, notice that some factors in $f_t$ may be observable. Observed factors such as the intercept, time effects, seasonal dummies, and other unit-invariant variables like index stock returns, central bank interest rates, and oil prices should be included in $X_i$. Any factor included in $X_i$ does not count towards $m$. 

Assumption \ref{ass:regressors} is similar to Assumption 2 of \citet{pesaran2006estimation} and states that the regressors must be stationary and that the cross-sectional dependence across units is fully captured by the factor structure. Assumption \ref{ass:errors} assumes a zero mean for errors, while it also allows for serial correlation, similar to \citet{pesaran2006estimation}. The errors must also be stationary. In combination with \ref{ass:commonfactors}, it also assumes that $x_{i,t}$ are strictly exogenous to $\varepsilon_{i,t}$; and because $q_{i,t}$ belongs to $x_{i,t}$, $q_{i,t}$ must also be exogenous. Yet $x_{i,t}$ can be weakly exogenous with respect to $e_{i,t}$, with feedback effects driven by $f_t$. Lags of the dependent variable are not allowed in $x_{i,t}$, as this would require lags of the cross-section averages of $y_{i,t}$ when estimating the factor space, but as we have argued, neither $\bar y_{t}$ nor its lags are permitted in our nonlinear setting. However, this does not become a restriction on the data generating process, as general forms of serial correlation are still permitted through both factors and errors.  

Assumption \ref{ass:contthreshvar}, states that the threshold variable has a continuous conditional probability density function and is uniformly bounded; see \citep{hansen2000sample}. Furthermore, it also excludes the possibility that $q_{i,t}=\gamma_i^0$ for all $t$. Assumption \ref{ass:identification} is a no perfect multicollinearity assumption allowing for identification of slopes and thresholds. It is a high level assumption, similar to those used elsewhere in the literature such as in \citet{miao2020panel1} and \citet{pesaran2006estimation}. It cannot be decomposed to more primitive assumptions, at least without significant restrictions on the data-generating process. Finally, Assumption \ref{ass:identification} also requires that there are observations on both regimes for each unit $i$. Assumption \ref{ass:threshspace} imposes the requirement that the threshold parameters belong to compact sets, as is standard in the literature. Assumption \ref{ass:fullrankH} contains full rank conditions that ensure matrix invertibility and Assumption \ref{ass:uniboundmom} imposes uniformly bounded moments as in \citet{kapetanios2011panels}, which are stronger than necessary but kept for simplicity. 

Assumption \ref{ass:shrink_thres} is called the ``shrinking threshold'' assumption and was introduced in the threshold literature in \citet{hansen2000sample}. This assumption is similar to the idea of local-to-zero approximations in hypothesis testing and is universally used in the threshold literature to derive a pivotal asymptotic distribution of $\hat{\gamma}$ such that tabulated critical values can be used. However, it also implies that the derived asymptotic distribution is a better approximation of the sampling distribution when $\delta^{0}_i$ is small, which nevertheless is the most useful case, since for large threshold magnitudes, $\delta_i^0$ becomes easier to estimate (see Theorem \ref{th:conv_gamma} below). This is the first paper in which the $\delta^{0}_i$ are allowed to have varying degrees of threshold magnitudes through the parameters $\alpha_i$. 

Part i) of Assumption \ref{ass:hiordmom} requires that the fourth-order conditional moments of $x_{i,t}$ and $x_{i,t}e_{i,t}$ exist and are bounded. Parts ii) and iii) are similar to Assumption A.5 of \citet{miao2020panel1}, and are made to ensure that the square matrix, $M_{\mathscr{D},i}(\gamma_i)$, is well-behaved in the neighbourhood of $\gamma_i^0$. Condition $M_{\mathscr{D},i}(\gamma^0_i) > 0$ excludes the continuous threshold model, similar to Assumption 1.7 in \citet{hansen2000sample}. $N \rightarrow \infty$ is not required here because $N \rightarrow \infty$ for the full heterogeneous model is only necessary to remove the IFE. Assumption \ref{ass:wellbehlim} provides conditional moment boundedness for all $\gamma_i \in \Gamma_i$.

\section{Asymptotic Theory}\label{section:asymptotictheory}
In this section, we derive the asymptotic theory by letting $N$ and $T$ tend to infinity. Theorems \ref{th:cons_gamma} and \ref{th:conv_gamma} prove the consistency of $\hat{\gamma}_i$ and derive its rate of converge. Theorems \ref{th:hetero_dis_thetai} and \ref{th:dis_gamma} derive the asymptotic distributions of the individual $\tilde{\theta}_{i}$ and $\hat{\gamma}_i$, while Theorem \ref{th:lr_thm} provides the likelihood ratio statistic to test hypotheses about $\gamma_i^0$. 

\begin{theorem} \label{th:cons_gamma}
		Under Assumptions \ref{ass:commonfactors}-\ref{ass:uniboundmom}, and as $(N,T) \rightarrow \infty,$
		    $\hat{\gamma_i} \stackrel{p}{\rightarrow} \gamma_i^0.$ 
\end{theorem}
Theorem \ref{th:cons_gamma} establishes the consistency for each $\hat{\gamma}_i$ under conditions weaker than \citet{miao2020panel1}, in that it does not require the shrinking threshold assumption. It also doesn't require any restrictions on the relative rate of expansion of $N$ and $T$, which is a property of CCE estimators \citep{pesaran2006estimation}. 

\begin{theorem} \label{th:conv_gamma}
		Under Assumptions \ref{ass:commonfactors}-\ref{ass:hiordmom}, $T^{\alpha_i}/\sqrt{N} \rightarrow 0$, as $(N,T) \rightarrow \infty$, $T^{1-2\alpha_i}(\hat{\gamma_i} - \gamma_i^0) = O_p(1)$.
\end{theorem}

Theorem \ref{th:conv_gamma} shows that the rate of convergence of $\hat{\gamma}_i$ is $T^{1-2\alpha_i}$, thus depending on the threshold effect magnitudes. Smaller $\alpha_i$ imply that the threshold effect is larger and in turn the convergence rate is faster. On the other hand, if $\alpha_i$ is close to $1/2$, the rate of convergence becomes slower since the magnitude of the threshold is smaller. A fixed-magnitude threshold effect is equivalent to $\alpha_i \rightarrow 0$ and results in $\hat{\gamma}_i - \gamma_i^0 = O_p[({T})^{-1}]$. The CCE literature typically requires $\sqrt{T}/N \rightarrow 0$ for individual unit estimators \citep{pesaran2006estimation}. Here we only require $T^{\alpha_i}/\sqrt{N} \rightarrow 0$ and thus, the relative rate of divergence between $T$ and $N$ in Theorem \ref{th:conv_gamma} is weaker than what is necessary elsewhere. The following two theorems derive the asymptotic distributions of the slope and threshold parameter estimators.

\begin{theorem} \label{th:hetero_dis_thetai} Under Assumptions \ref{ass:commonfactors}-\ref{ass:hiordmom} and conditional on $\mathscr{D}$,  as $(N,T) \rightarrow \infty$ with $T^{\alpha_i}/\sqrt{N} \to 0$ and $\sqrt{T}/N \rightarrow 0$, then $\sqrt{T}\tilde{V}_{\theta_i}^{-1/2}(\tilde{\theta}_i - \theta^0_{i}) \hspace{0.1cm} \stackrel{d}{\rightarrow} \hspace{0.1cm} N(0,I_{K+r})$.  
\end{theorem}

\begin{theorem} \label{th:dis_gamma}
	Under Assumptions \ref{ass:commonfactors}-\ref{ass:wellbehlim},
  if $T^{\alpha_i}/\sqrt{N} \rightarrow 0$, and $\sqrt{T}/N \rightarrow 0$ as $(N,T) \rightarrow \infty$,  
   then:
		\begin{equation}
		T^{1-2\alpha_i}(\hat{\gamma}_i - \gamma^0_i) \hspace{0.1cm} \stackrel{d}{\rightarrow} \hspace{0.1cm} \phi_i {\zeta},
		\end{equation}
		where $\phi_i = V_i(\gamma^0_i)/D_i(\gamma^0_i)^2$,  $\zeta = \mathop{argmax}_{-\infty < s < \infty} \{(-1/2)|s|+ B(s) \}$, and $B(s)$ is a two-sided standard Brownian motion on the real line.
	\end{theorem}

The asymptotic distribution of the threshold parameters is the same as in \cite{hansen2000sample}. In the presence of conditional homoskedasticity in $\varepsilon_{i,t}$, $\phi_i$ further simplifies to $\sigma_{\varepsilon,i}^2[D_i(\gamma^0_i)]^{-1}$. Serial correlation nuisance parameters do not enter the asymptotic distribution of $\hat\gamma_i$, because its variance, as can be seen from $V_{i,T}$ and $D_{i,T}$, depends not only on $\varepsilon_{i,t}$, but also on the distribution of the threshold variable $q_{i,t}$. In the proof of the theorem, we show that the joint probability that both $q_{i,t}$ and $q_{i,t'}$ for $t\neq t'$ appear in a small neighborhood of $\gamma_i^0$ is asymptotically 0, because the size of that neighborhood is shrinking due to the shrinking threshold assumption \ref{ass:shrink_thres}. Therefore, serial correlation cross-terms with $t$ and $t'$ are asymptotically negligible.

\citet{hansen2000sample} suggests that the asymptotic distribution in Theorem \ref{th:dis_gamma} should not be used to build confidence intervals for $\gamma_i$, because $\phi_i$ is difficult to estimate accurately. We therefore propose a likelihood ratio test for the null hypothesis $H_0:\gamma_i^0=\gamma_i$ based on:
\begin{align*} 
	    LR(\gamma_i)= \left[RSS(\hat{\theta}_i(\gamma_i),\gamma_i) - RSS(\tilde{\theta}_i,\hat{\gamma}_i)\right]/\hat{\sigma}_{\varepsilon,i}^2,
	\end{align*}
	where $RSS(\hat{\theta}_i(\gamma_i),\gamma_i) = (\Tilde{y}_i - \Tilde{Z}_i(\gamma_i)\hat{\theta}_i(\gamma_i))'(\Tilde{y}_i - \Tilde{Z}_i(\gamma_i)\hat{\theta}_i(\gamma_i))$, and $\hat{\sigma}_{\varepsilon,i}^2 = T^{-1}RSS(\tilde{\theta}_i,\hat{\gamma}_i)$.
    \begin{theorem} \label{th:lr_thm}
		Under $H_0: \gamma^0_i = \gamma_i$, Assumptions \ref{ass:commonfactors}-\ref{ass:wellbehlim}, as $(N,T) \rightarrow \infty$ with $T^{\alpha_i}/\sqrt{N} \rightarrow 0$ and $\sqrt{T}/N \rightarrow 0$: 
		$$LR_i(\gamma^0_i) \stackrel{d}{\rightarrow} \hspace{0.1cm} \eta_i^2 \Xi,$$
		where $\eta_i^2 = V_i(\gamma_i^0)/\sigma^{\star2}_{\varepsilon,i}D_i(\gamma^0_i)$, $\sigma^{\star2}_{\varepsilon,i} = plim_{T \rightarrow \infty} \hat{\sigma}_{\varepsilon,i}^2$. The random variable $\Xi = \mathop{sup}_{s \in \mathbb{R}}[2B(s) - |s|]$ has a distribution function given by $\mathbb{P}( \Xi \leq x) = (1-exp(-x/2))^2$. An estimator for $\eta_i$ is presented in the supplementary online appendix.
	\end{theorem}

In a special case where $\varepsilon_{i,t}$ is homoskedastic, $\eta_i^2 = 1$ and inference can be made based on readily available critical values. The inverted distribution function of $\Xi$ is given by $c(a) = -2log(1-\sqrt{1-a})$, where $c(a)$ is the critical value and $a$ is the significance level. For various $a$'s the $c(a)$ can be found in Table 1 of \citet{hansen2000sample}, i.e., for $a=0.1$ it is $5.94$, for $a=0.05$ it is $7.35$ and for $a=0.01$ it is $10.59$. These critical values are used to test the null hypothesis $H_0:\gamma_i=\gamma_i^0$ with rejection region $LR_i(\gamma_i^0) > c(\alpha)$. 
Confidence sets for $\gamma_i$ can thus be obtained by inverting this family of likelihood-ratio tests. The asymptotic confidence set with coverage probability $1-\alpha$ is defined as $\left\{ \gamma_i : LR_i(\gamma_i) \le c(\alpha) \right\}$. In practice, this set is obtained by plotting $LR_i(\gamma)$ as a function of $\gamma$ and identifying the values of $\gamma$ for which the statistic lies below the horizontal line at $c(\alpha)$.

Theorem \ref{th:lr_thm} is asymptotically correct under the shrinking threshold assumption $\delta^{0}_i \rightarrow 0$ for all $i = 1,2,...,N$. However, if the errors are homoskedastic and also normal and i.i.d., we hypothesize that the intuition of Theorem 3 of \citet{hansen2000sample} holds here as well, and therefore inference based on the LR test is asymptotically valid even if $\delta^0_i$ does not shrink.

\section{A Semi-Homogeneous Threshold Regression Model}
In the previous fully heterogeneous model we used cross-sectional data only to eliminate the unobserved IFE. However, a large cross-section dimension can bring additional benefits when some parameters are assumed to be the same across units. These benefits include faster convergence rates and more efficient estimation. Here we consider a special case of \eqref{basicmodel} with $\gamma_i=\gamma$: 
\begin{equation} \label{homobasicmodel}
	y_{i,t} = \beta_{i}'x_{i,t} + \delta_{i}'w_{i,t} \mathbb{I}\{q_{i,t} \leq \gamma \}+ e_{i,t}.
\end{equation}
This model lies in between the fully heterogeneous model \eqref{basicmodel} and the fully homogeneous model of \cite{miao2020panel1}. We call it ``semi-homogeneous'' and in the supplementary appendix we include a BIC-type criterion that can choose between models \eqref{basicmodel} and \eqref{homobasicmodel}. 

The combination of a common $\gamma$ and unit-specific $\beta_i$ and $\delta_i$ creates complications for parameter identification. The identification assumption \ref{ass:identification} implies that all units need enough time series observations in both regimes; explicitly $\sum_{t=1}^{T} \mathbb{I}\{q_{i,t} \leq \gamma\} > 0$ and $\sum_{t=1}^{T} \mathbb{I}\{q_{i,t} > \gamma\} > 0$, for every unit $i$. However, if the supports of $q_{i,t}$ and $q_{j,t}$, for $i \neq j$, are disjoint, then the aforementioned condition will not hold and either $\delta_i$ or $\delta_j$ will not be identified. Consider, for example, the impact of government expenditure on economic growth. The UK's General government final consumption expenditure (\% of GDP) varies between 16\% to 22\% from 1973 to 2021, while that of Mexico in the same period varies between 8\% to 12\%. Therefore, there is no common $\gamma$ that creates two regimes in both countries, and hence one of the $\delta_i$'s will not be identified.

The discussion above makes clear that the semi-homogeneous model is applicable only when the threshold variable has a common support across $i$. Despite this limitation, there are many applications where this happens, such as when the threshold variable results from transformations such as quantile, standardisation, scaling, or ratio, among others. For example, \citet{nocera2023causal} study the Fed's large-scale asset purchases on firms' capital structure and use as a threshold variable the quantile transformation of the ratio of debt to assets. \citet{girma2005absorptive} studies the non-linear impact of absorptive capacity by constructing the threshold variable as $q_{i,t}/q^{\star}_{i,t}$,  where $q^{\star}_{i,t}$ is the maximum level of $q_{i,t}$ and $q_{i,t} > 0$, such that the threshold variable shares a common support between $0$ and $1$. Yet another alternative would be to assume $\delta_i=\delta$, as in \cite{chudik2017there}. If correct, this last assumption removes the identification problem and even leads to a pooled $\tilde\delta$ estimator with a faster rate of convergence. The estimator is given in \eqref{delta_p}, in the next section.

In the semi-homogeneous model, interest shifts to $\gamma$ and the means of the individual coefficients, $\theta$. Assuming $\gamma$ is known, the pooled Mean Group (MG) CCE estimators are:
\begin{align}
        \hat{\theta}(\gamma) = \frac{1}{N} \sum_{i=1}^{N} \hat{\theta}_i(\gamma),  & \;\;\; \text{where} \;\;\;
     \hat{\theta}_i (\gamma)= \left(\tilde{Z}_i (\gamma)' \tilde{Z}_i (\gamma) \right)^{-1} \tilde{Z}_i (\gamma)' \tilde{y}_i. \label{MG_est}
\end{align}  

If $\gamma$ is unknown, we estimate it as the $\gamma$, which minimises the CCE MG sum of squared residuals: $ \hat{\gamma} =\text{argmin}_{\gamma \in \Gamma} \sum_{i=1}^{N}\left[\tilde{y}_i- \tilde{Z}_i (\gamma) \hat{\theta}_i(\gamma) \right]'\left[\tilde{y}_i - \tilde{Z}_i (\gamma) \hat{\theta}_i(\gamma)\right]$. The latter is a step function for $\gamma$ and can be trimmed as described in Section \ref{section:estimation}. In the following, we denote $\tilde{\beta}_i = \hat{\beta}_i (\hat{\gamma})$, $\tilde{\delta}_i = \hat{\delta}_i (\hat{\gamma})$, $\tilde{\beta} = \hat{\beta}(\hat{\gamma})$, $\tilde{\delta} = \hat{\delta} (\hat{\gamma})$, $\tilde{\theta}_i = (\tilde{\beta}_i , \tilde{\delta}_i)$ and $\tilde{\theta} = (\tilde{\beta} , \tilde{\delta})$. The estimator variances are given by $\tilde{V}_{\theta_i}= \hat{V}_{\theta_i}(\hat\gamma)=\hat{\Sigma}_i^{-1}(\hat\gamma)\hat{S}_{i}(\hat\gamma)\hat{\Sigma}_i^{-1}(\hat\gamma)$, and $\tilde{V}_{\theta}=(N-1)^{-1}\sum_{i=1}^N(\tilde\theta_i-\tilde\theta)(\tilde\theta_i-\tilde\theta)'$.  To proceed we need to make the following assumptions which apply only to the semi-homogeneous model.  

\setcounter{assumption}{0}
\renewcommand{\theassumption}{S.\arabic{assumption}}

\begin{assumption}[Distribution of Heterogeneity]\label{ass:p_obs_het}
The slopes $\theta_{i}=(\beta_i',\delta_i')'$ follow the random coefficient model $\theta^0_{i} = \theta^0 + v_{i}$, with $v_{i} \sim i.i.d. (0,\Sigma_{v})$ for $i = 1,2,...,N$,
where $ || \theta^0 || < C $. $\Sigma_v$ is a $(K+r) \times (K+r)$ symmetric non-negative definite matrix such that $|| \Sigma_v || < C$, and $v_i$ are distributed independently of $\lambda_j$, $\Pi_j$, $\xi_{j,t}$, $\varepsilon_{j,t}$ and $f_t$ $\forall \; i$, $j$ and $t$.  
\end{assumption}

\begin{assumption}[Pooled Shrinking Threshold Assumption] \label{ass:p_shrink_thres}
		Let $C_{0,i} = C_0 + C_{v_i}$, where $C_0$ is fixed with $C_0 \in \mathbb{R}$, and  $C_{v_i} \sim iid(0,\Sigma_{v_i})$, where $\Sigma_{v_i}=L'\Sigma_{v}L$, where $L=[0_{r\times K},  I_r]'$. Furthermore, $sup_{i} ||C_{0,i}|| < \infty$. For $0 < \alpha < 1/2$, the threshold effect $\delta^0_{i}$ satisfies $\delta^0_{i} = C_{0,i}T^{-\alpha}$ for $i = 1,2,...,N$, and therefore, by Assumption \ref{ass:p_obs_het}, $\delta^0 = C_0T^{-\alpha}$.   
\end{assumption}

\begin{assumption}[Threshold Parameter Space]\label{ass:p_thres_space} The threshold parameter $\gamma^0\in \Gamma=[\underline{\gamma},\overline{\gamma}]$, where $\Gamma$ is a compact set.
\end{assumption}

\begin{assumption}[Full Rank Conditions] \label{ass:p_full_rank}
Let $M_{NT}^{\star} = E_{\mathscr{D}}\left[(NT)^{-1}\sum_{i=1}^N {X}_i'M_FX_i\right]$, and 
$M_{NT}^{\star}(\gamma^1,\gamma^2) = E_{\mathscr{D}}\left[(NT)^{-1}\sum_{i=1}^N X_i(\gamma^1)'M_FX_i(\gamma^2)\right]$, where for $\gamma^1 = \gamma^2 = \gamma$, $M_{NT}^{\star}(\gamma)=M_{NT}^{\star}(\gamma^1,\gamma^2)$. Define also $M^{\star\star}_{NT}(\gamma) = E_{\mathscr{D}}\left[(NT)^{-1}\sum_{i=1}^N X'_iM_FX_i(\gamma)\right]$, all conditionally on $\mathscr{D}$. Then, we require that i)	$M_{T,i}^{\star}$, and $ M_{T,i}^{\star}(\gamma^1,\gamma^2) > 0$ for all $i$, and $\gamma^1, \gamma^2 \in \Gamma$, and ii) for all $\gamma\in\Gamma$:  
\begin{equation}
    Q_{1}(\gamma) = \begin{bmatrix}
		M_{NT}^{\star} & M_{NT}^{\star\star}(\gamma)\\
		M_{NT}^{\star\star}(\gamma) & M_{NT}^{\star}(\gamma)\\
		\end{bmatrix} > 0.
\end{equation}
\end{assumption} 

\begin{assumption}[Identification Condition] \label{ass:identification_semi}   
i) $\mathcal{S}_{1,i}(\gamma) = \Tilde{Z}_i(\gamma)'\Tilde{Z}_i(\gamma)>0$ for all $i$, ii) $N^{-1}\sum_{i = 1}^{N}||\delta_i||^2 \neq 0$, and 
iii) for $\tau > 0$, as $T  \rightarrow \infty$, and 
for all $i$ uniformly, it holds that:  $\mathbb{P}\left\{\min_{\gamma \in \Gamma} \lambda_{min}\left[T^{-1}\left(\mathcal{S}_{3,i}(\gamma) - \mathcal{S}_{2,i}(\gamma)'\mathcal{S}_{1,i}^{-1}(\gamma)\mathcal{S}_{2,i}(\gamma)\right)\right] \geq \tau min \left(1, |\gamma - \gamma^0|\right) \right\} \stackrel{p}{\rightarrow} 1$.
\end{assumption}

\begin{assumption}[Higher Order Moments]\label{ass:p_hi_ord}
	Let $M_{\mathscr{D}}(\gamma)=N^{-1}\sum_{i=1}^NM_{\mathscr{D},i}(\gamma)$. 
    i) There exist $\mathscr{D}$-dependent variables $C_{i,t}^{\mathscr{D}}$ such that
    $sup_{\gamma \in \Gamma} E_{\mathscr{D}}(||x_{i,t} ||^4 |q_{i,t} = \gamma) \leq C_{i,t}^{\mathscr{D}}$,  
     $sup_{\gamma \in \Gamma} E_{\mathscr{D}}(||x_{i,t}e_{i,t} ||^4 |q_{i,t} = \gamma) \leq C_{i,t}^{\mathscr{D}}$, and $\mathbb{P}\Big((NT)^{-1}\sum_{i=1}^{N}\sum_{t=1}^{T}{C_{i,t}^{\mathscr{D}}}^2 \leq C^2 \Big) = 1$, for some $C < \infty$, as $(N,T) \rightarrow \infty$. ii) $M_{\mathscr{D}}(\gamma^0) > 0$ a.s., and $M_{\mathscr{D}}(\gamma)$ is continuous at $\gamma = \gamma^0$. iii) For all $\epsilon > 0$, there exist constants $B > 0$ and $C_1 > 0$, for sufficiently large $N,T$, we have $ \mathbb{P}\Big(\mathop{inf}_{|\gamma - \gamma^0| < B} M_{\mathscr{D}}(\gamma) > C_1 \Big) > 1 - \epsilon.$
\end{assumption}

\begin{assumption} [Well-behaved Limits] \label{ass:p_well_beh_lim} Let $D(\gamma) = plim_{(N,T) \rightarrow \infty}D_{NT}(\gamma)$ and $V(\gamma) = plim_{(N,T) \rightarrow \infty}V_{NT}(\gamma)$, where $ D_{NT}(\gamma) = (NT)^{-1}\sum_{i=1}^{N}\sum_{t=1}^{T}E\left[(C_{0,i}'x_{i,t})^2|q_{i,t} = \gamma\right]f_{i,t}(\gamma)$ and $V_{NT}(\gamma) = (NT)^{-1}\sum_{i=1}^{N}\sum_{t=1}^{T}E\left[(C_{0,i}'x_{i,t})^2\varepsilon_{i,t}^2 |q_{i,t} = \gamma\right]f_{i,t}(\gamma)$. Then, i) the limits $D(\gamma)$ and $V(\gamma)$ exist, are continuous at $\gamma = \gamma^0$, and $D(\gamma^0) > 0, V(\gamma^0) > 0$. ii) There is a constant $C^{\star}$ such that: $Var\left[(NT)^{-1}\sum_{i=1}^{N}\sum_{t=1}^{T}E_{\mathscr{D}}(||g^{(1)}_{i,t}(\gamma^1,\gamma^2) ||^2)  \right]  \leq C^\star sup_{i,t} E(||g^{(1)}_{i,t}(\gamma^1,\gamma^2) ||^4)$ and $Var\left[(NT)^{-1}\sum_{i=1}^{N}\sum_{t=1}^{T}E_{\mathscr{D}}(||g^{(2)}_{i,t}(\gamma^1,\gamma^2) ||^2) \right] \leq C^\star sup_{i,t} E(||g^{(2)}_{i,t}(\gamma^1,\gamma^2) ||^4)$.
\end{assumption}

Assumption \ref{ass:p_obs_het} imposes $i.i.d.$ distributed heterogeneous coefficients which are randomly distributed across units and independent of any other random elements in the model, as in Pesaran (2006). This assumption allows the use of the pooled MG CCE estimator for $\theta$. If \ref{ass:p_obs_het} fails, so does the consistency of the MG estimators. Assumption \ref{ass:p_shrink_thres} is a variant of the shrinking threshold assumption \ref{ass:shrink_thres} where the shrinking parameter $\alpha$ is now common across units. Notice that since $\delta_i$ are heterogeneous, the rate of shrinkage is $O(T^{-\alpha})$, instead of $O[(NT)^{-\alpha}]$ as in \cite{miao2020panel1}. Assumptions \ref{ass:p_thres_space} - \ref{ass:p_well_beh_lim} are variations of \ref{ass:identification}, \ref{ass:threshspace},  \ref{ass:fullrankH}, \ref{ass:hiordmom} and \ref{ass:wellbehlim} respectively. 

For MG estimators, the scenario where $m$ is strictly smaller than $K$ is problematic, as having more cross-section averages than number of factors can induce bias to the CCE estimators. The same can happen in the presence of distinct factors, that is, when the factors in equation \eqref{or_eq7} are different from the factors in \eqref{or_eq8}. These issues are dealt with regularization and bootstrap in \cite{juodis2022} and \cite{DeVosStauskas2024}, and the same solutions are applicable in our setting. The basic intuition for why the regularization and the bootstraps should work in our non-linear threshold model is because, if the thresholds are known, the threshold regression model becomes linear, and hence falls within the class of models studied in those two papers. If the thresholds are unknown however, they can be estimated consistently as the CCE estimators remain consistent. Therefore, asymptotically observing $\hat \gamma_i$ is just as good as observing $\gamma_i $ itself.  On a separate topic, the above intuition can also be used to justify using the test of \cite{JuodisReese2022} to test for remaining cross-section dependence in the threshold regression residuals. This intuition is supported by Monte Carlo results in the supplementary online appendix.

As previously, lagged dependent variables are not allowed and should be left as serial correlation in the errors. This approach has a double benefit; first, it allows the use of the bootstrap in \cite{DeVosStauskas2024} extending the applicability of the new methods to the cases mentioned earlier, and second, it does not require bias correction for the removal of the $O(T^{-1})$ Nickell bias which would be introduced by lags of $y_{i,t}$. This bias can in theory be removed by the half-panel jacknife estimator \citep{juodisreese2026} but it is not entirely clear how this is done in non-linear panels and it will certainly require further assumptions on the regime prevalence in the sub-panel estimations, notwithstanding of course the restriction that $m=1$ due to the threshold nonlinearity.
Theorem \ref{th:gamma_pool} repeats the main results for a pooled $\hat \gamma$, while Theorem \ref{dis_thetamg} derives the rate of convergence and the asymptotic distribution of the MG estimator $\tilde\theta$:  

\begin{theorem} 
\label{th:gamma_pool}
The following results hold:
i) Under the assumptions \ref{ass:commonfactors}-\ref{ass:contthreshvar}, \ref{ass:uniboundmom}, \ref{ass:p_obs_het}, \ref{ass:p_thres_space}-S.5, and as $(N,T) \rightarrow \infty,$
		    $\hat{\gamma} \stackrel{p}{\rightarrow} \gamma^0.$
            
ii) Under the assumptions \ref{ass:commonfactors}- \ref{ass:contthreshvar},  \ref{ass:uniboundmom}, and \ref{ass:p_obs_het}-\ref{ass:p_hi_ord}, $T^{\alpha}/\sqrt{N} \rightarrow 0$, as $(N,T) \rightarrow \infty$, $NT^{1-2\alpha}(\hat{\gamma} - \gamma^0) = O_p(1)$.

iii) Under the assumptions \ref{ass:commonfactors}- \ref{ass:contthreshvar},  \ref{ass:uniboundmom}, and \ref{ass:p_obs_het}-\ref{ass:p_hi_ord},  if $T^{\alpha}/\sqrt{N} \to 0$, $\sqrt{T}/N \rightarrow 0$, and conditional on $\mathscr{D}$, then the limiting distribution is $\sqrt{T}\tilde{V}_{{\theta}_i}^{-1/2}(\tilde{\theta}_i - \theta^0_{i}) \hspace{0.1cm} \stackrel{d}{\rightarrow} \hspace{0.1cm} N(0,I_{K+r})$. 
iv) Under the assumptions \ref{ass:commonfactors}- \ref{ass:contthreshvar},  \ref{ass:uniboundmom}, and \ref{ass:p_obs_het}-\ref{ass:p_well_beh_lim}, let $\phi = V(\gamma^0)/D(\gamma^0)^2$. If in addition $T^{\alpha}/\sqrt{N} \rightarrow 0$, and $\sqrt{T}/N \rightarrow 0$ as $(N,T) \rightarrow \infty$, then:
		\begin{equation}
		NT^{1-2\alpha}(\hat{\gamma} - \gamma^0) \hspace{0.1cm} \stackrel{d}{\rightarrow} \hspace{0.1cm} \phi \zeta
\end{equation}
\end{theorem}

\begin{theorem} \label{dis_thetamg}
		Under assumptions \ref{ass:commonfactors}-\ref{ass:contthreshvar},  \ref{ass:uniboundmom}, and \ref{ass:p_obs_het}-\ref{ass:p_hi_ord}, as $(N,T) \rightarrow \infty$ and $T^{\alpha}/\sqrt{N} \rightarrow 0
		$, we have: 

i)
  \begin{equation}\label{eq_thetabar_dist}
      \begin{bmatrix}
		\sqrt{N} I_K& 0_{K\times r} \\
		0_{r\times K} &\sqrt{N}T^{\alpha} I_{r}
		\end{bmatrix}(\tilde{\theta} - \theta^0) \hspace{0.1cm} \stackrel{d}{\rightarrow} \hspace{0.1cm} N(0,\Sigma_v),
  \end{equation} 

ii) $N(\tilde{\theta} - \theta^0)'\tilde{V}_{\theta}^{-1}(\tilde{\theta} - \theta^0) \hspace{0.1cm} \stackrel{d}{\rightarrow} \hspace{0.1cm} \chi^2_{K+r}$.
\end{theorem}

Theorem \ref{dis_thetamg} demonstrates that the heterogeneity of the slope coefficient has unique implications for the asymptotic theory, as $\tilde{\delta}$ has a rate of convergence that is much faster than that of the MG estimator in linear regression, see, e.g., \citet{JKS}. This arises due to the shrinking threshold assumption \ref{ass:shrink_thres}:
\begin{equation}
    \delta^0_i=\frac{C_{0,i}}{T^\alpha}=\frac{C_{0}}{T^\alpha}+\frac{C_{v_i}}{T^\alpha},
\end{equation}
which implies that the error term which drives unit heterogeneity $C_{v_i}/T^\alpha$, is $O(T^{-\alpha})$. Hence, in the limit, heterogeneity vanishes and we have a homogeneous model. The closer the $\delta_i^0$ are to zero, the closer they are to each other, and thus the shrinking threshold effects are becoming ever more homogeneous. This is reflected in the estimator rate of convergence. Large threshold effects correspond to $\alpha \to 0$, yielding the standard MG convergence rate $\sqrt{N}$. In contrast, when threshold effects are small ($\alpha \to 1/2$), the rate improves to $\sqrt{NT}$ in the limit, which is the standard pooled rate found in homogeneous models \citep{pesaran2006estimation}. Part i) of Theorem \ref{dis_thetamg} cannot be used in practice because it contains the unknown $\alpha$. Hypothesis tests are facilitated by part ii), which is due to the self-normalization of the variance estimator. 

The following theorem  develops a test for testing hypotheses on the threshold parameter.
\begin{theorem} \label{lr_thm}
Under $H_0: \gamma = \gamma^0$, assumptions \ref{ass:commonfactors}-\ref{ass:contthreshvar}, \ref{ass:uniboundmom} and \ref{ass:p_obs_het}-\ref{ass:p_well_beh_lim}, $T^{\alpha}/\sqrt{N} \rightarrow 0$, and $\sqrt{T}/N \rightarrow 0$, as $(N,T) \rightarrow \infty$, it holds that
$LR(\gamma^0) \stackrel{d}{\rightarrow} \hspace{0.1cm} \eta^2 \Xi$. In this case, $LR(\gamma)= \left[RSS(\hat{\theta}_i(\gamma),\gamma) - RSS(\hat{\theta}_i(\hat{\gamma}),\hat{\gamma})\right]/\hat{\sigma}_{\varepsilon}^2$, and $\eta^2 = V(\gamma^0)/\sigma^{\star2}_{\varepsilon}D(\gamma^0)$, with $\sigma^{\star2}_{\varepsilon} = plim_{(N,T)\rightarrow \infty} \hat{\sigma}_{\varepsilon}^2$. Additionally,  $RSS(\hat{\theta}_i(\gamma),\gamma)=\sum_{i=1}^{N}\left[\tilde{y}_i- \tilde{Z} (\gamma) \hat{\theta}_i(\gamma) \right]'\left[\tilde{y}_i - \tilde{Z}_i (\gamma) \hat{\theta}_i(\gamma) \right]$, and the error variance estimator is $\hat{\sigma}_{\varepsilon}^2 = (NT)^{-1}RSS(\hat{\theta}_i(\hat{\gamma}),\hat{\gamma})$.
\end{theorem} 

The parameter $\eta^2=1$ if the errors are homoskedastic across both $N$ and $T$. If not, a nonparametric estimator is given in the supplementary online appendix.

\section{Testing the Null Hypothesis of No Threshold Regression}

In practice it may not be known if there is non-linearity in the model, or which is the threshold variable $q_{i,t}$. In Hansen (2000), both questions are addressed by applying a test of linearity (no threshold regression) to a known $q_{i,t}$ in the first case, or for each potential threshold variable in the second case. In this section, we enable the same type of inference by providing tests of linearity for both the fully heterogeneous and the semi-homogeneous models. In the first model, for each $i$ the null hypothesis is $H_0:\delta_i= 0$ and the alternative is $H_1:\delta_i\neq 0$. In the second model, there is a joint null hypothesis $H_0:\delta_i= 0$ for all $i$ against the alternative that at least one $\delta_i\neq 0$ over all $i$. 

Testing the null hypothesis of no threshold regression is challenging because under $H_0: \delta^0_i = 0$, for any model, the threshold parameters disappear and thus cannot be identified. To deal with this problem, we follow \citet{hansen1996inference} and test the null hypothesis based on a supremum-type Wald statistic whose limiting distribution is non-standard but can be approximated by the bootstrap. In the fully heterogeneous model, for unit $i$ we employ:
\begin{align}
supW_i &= \mathop{sup}_{\gamma_i \in \Gamma_i}\mathcal{W}_{i}(\gamma_i),\; \text{where} \; \mathcal{W}_{i}(\gamma_i) = T\hat{\delta}_i(\gamma_i)'\hat{V}^{-1}_{\delta_i}(\gamma_i)\hat{\delta}_i(\gamma_i),
\end{align}
where $\hat{V}_{\delta,i}(\gamma_i) = L'\hat{V}_{\theta_i}(\gamma_i) L$, and $L$ is the selection matrix defined in Assumption \ref{ass:p_shrink_thres}.
To derive the asymptotic distribution of $supW_i$ we will need an additional assumption:

\setcounter{assumption}{0}
\renewcommand{\theassumption}{NLH.\arabic{assumption}}

\begin{assumption}\label{ass:NLH}
Define $S_{i,T}(\gamma_i) = \sqrt{T}^{-1}Z_i(\gamma_i)'M_{F}\varepsilon_i$, and additionally,
$\Sigma_i(\gamma^1_i,\gamma^2_i) = plim_{T \rightarrow \infty}T^{-1} E(Z_i(\gamma^1_i)'M_{F}Z_i(\gamma^2_i)|\mathscr{D})$. Then, as $T\to\infty$,     $S_{i,T}(\gamma_i) \Rightarrow S_i(\gamma_i)$, where $S_i(\gamma_i)$ is a mean-zero Gaussian process with kernel $K_i(\gamma^1_i,\gamma^2_i) = plim_{T \rightarrow \infty}T^{-1}E(Z_i(\gamma^1_i)'M_{F}\Sigma_{\varepsilon_i}M_{F}Z_i(\gamma^2_i)|\mathscr{D})$, and $\Sigma_{\varepsilon_i} = E(\varepsilon_i\varepsilon_i')$.
\end{assumption}

This is a high-level assumption that is straightforward to justify. More primitive conditions can be found in \citet{hansen1996inference} or in Lemma A.9 of \citet{miao2020panel1}.

\begin{theorem} \label{thm_mgtest} Suppose that Assumptions \ref{ass:commonfactors}-\ref{ass:contthreshvar}, \ref{ass:threshspace}-\ref{ass:uniboundmom} and \ref{ass:NLH} hold, as $(N,T) \rightarrow \infty$, under $H_{0}: \delta^0_i = 0$ for a given $i$, then, $supW_i \stackrel{d}{\rightarrow}  \mathop{sup}_{\gamma_i \in \Gamma_i}\mathcal{W}^c_{i}(\gamma_i)$, where the random process $\mathcal{W}^c_{i}(\gamma_i) = \bar{S}_i(\gamma_i)'\Bar{K}_i(\gamma_i,\gamma_i)^{-1}\bar{S}_i(\gamma_i)$, and $\Bar{S}_i(\gamma_i) = L'\Sigma_i(\gamma_i,\gamma_i)^{-1}S_i(\gamma_i)$ is a mean-zero Gaussian process with covariance kernel $\Bar{K}_i(\gamma^1_i,\gamma^2_i) = L'\Sigma_i(\gamma^1_i,\gamma^1_i)^{-1}K_i(\gamma^1_i,\gamma^2_i)\Sigma_i(\gamma^2_i,\gamma^2_i)^{-1}L$.
\end{theorem}

Note that the above theorem does not require the shrinking threshold assumption \ref{ass:shrink_thres}, while $N \rightarrow \infty$ is only necessary to remove IFE and thus, there are no other restrictions on the relative rate of divergence between $N$ and $T$.

Moving on to the semi-homogeneous model, a new challenge arising due to heterogeneity is that the joint null hypothesis of no threshold regression is equivalent to testing $N$ individual hypotheses, with $N$ going to infinity. This is a multiple testing problem that can lead to low power. To avoid multiple testing, we employ the approach of \citet{JKS}, which exploits the fact that, under the null, the model becomes homogeneous in $\delta_i^0$ because $\delta_i^0=\delta^0=0$ for all $i$. The null implies a model with homogeneous $\delta$ coefficient, which for a given $\gamma$ can be estimated by the pooled CCE estimator:
\begin{equation}\label{delta_p}
\hat{\delta}_p(\gamma) = \left(\sum_{i=1}^{N}W_i(\gamma)'M_{Z^{\star}_i(\gamma)}W_i(\gamma)\right)^{-1}\left( \sum_{i=1}^{N}W_i(\gamma)'M_{Z^{\star}_i(\gamma)}y_i\right),
\end{equation}
where $Z^{\star}_i(\gamma) = (\bar{Z}(\gamma),X_i) = (\Bar{X},\Bar{W}(\gamma),X_i)$. The estimator $\hat{\delta}_p(\gamma)$ is a pooled estimator with a faster, $\sqrt{NT}$ rate of convergence. Note that $\hat{\delta}_p(\gamma)$ is different from $\hat{\delta}(\gamma)$ due to the annihilator matrix. In $M_{Z^{\star}_i(\gamma)}$, $\Bar{W}(\gamma)$ is included to remove asymptotic bias from the interactive effects following \citet{karavias2022structural}, and $X_i$ is included to project out the variables with heterogeneous coefficients as in \citet{JKS}. In the following, for any $T$-rowed matrix $A$, let $\tilde{A}(\gamma) = M_{\bar{Z}(\gamma)}A$. The supremum Wald statistic based on $\hat\delta_p(\gamma)$ is:
\begin{align}
supW &= \mathop{sup}_{\gamma \in \Gamma}\mathcal{W}_{NT}(\gamma),\; \text{where} \;  \mathcal{W}_{NT}(\gamma) = NT\hat{\delta}_p(\gamma)'\hat{V}^{-1}_{\delta_p}(\gamma)\hat{\delta}_p(\gamma).
\end{align}
$\hat{V}_{\delta_p}(\gamma) = L'\hat{\Sigma}^{-1}(\gamma)\hat{K}(\gamma)\hat{\Sigma}^{-1}(\gamma)L$, with $\hat{\Sigma}(\gamma) = (NT)^{-1}\sum_{i=1}^{N}\tilde{Z}_i(\gamma)'\tilde{Z}_i(\gamma)$, and variance $\hat{K}(\gamma) = N^{-1}\sum_{i=1}^{N}\hat{K}_i(\gamma)$, where $\hat{K}_i(\gamma)=\hat S_i(\gamma)$ defined in equation \eqref{NW_eq}, but where $\tilde{z}_{i,t}(\gamma)$ is now based on the $ M_{\bar{Z}(\gamma)}$ annihilator matrix, and the same applies to $\hat{\varepsilon}_i(\gamma)$.

The implementation of $supW$ requires an approximation of $\Gamma$ similar to the one used to estimate $\hat\gamma$ above. Theorem \ref{thm_mgtest} derives the limiting distribution of $supW$, based on a pooled version of Assumption \ref{ass:NLH}. However, it is additionally required that the mean of the $\delta_i$ is different from zero, because in this case the test would have no power due to the use of $\hat\delta_p(\gamma)$. This is because if $E(\delta_i^0)=\delta^0=0$, then the null $H_0:\delta^0=0$ is true. This assumption is not considered to be strong, as threshold effects are typically expected to move in the same direction across units rather than offset each other. 

\setcounter{assumption}{0}
\renewcommand{\theassumption}{NLP.\arabic{assumption}}

\begin{assumption}\label{ass:NLP} i) $\delta^0\neq0$. ii)
    Define $S_{NT}(\gamma) = (\sqrt{NT})^{-1}\sum_{i=1}^{N}Z_i(\gamma)'M_{F}\varepsilon_i$ and also $\Sigma(\gamma^1,\gamma^2) = plim_{(N,T) \rightarrow \infty}(NT)^{-1}\sum_{i=1}^{N} E(Z_i(\gamma^1)'M_{F}Z_i(\gamma^2)|\mathscr{D})$. Then, as $N,T\to\infty$, 
    $S_{NT}(\gamma) \Rightarrow S(\gamma)$, where $S(\gamma)$ is a zero-mean Gaussian process with covariance kernel given by $K(\gamma^1,\gamma^2) = plim_{N,T \rightarrow \infty}(NT)^{-1}$ $E(\sum_{i=1}^{N}Z_i(\gamma^1)'M_{F}\Sigma_{\varepsilon_i}M_{F}Z_i(\gamma^2)|\mathscr{D})$.
\end{assumption}

\begin{theorem} \label{thm_mgtest} Suppose that assumptions \ref{ass:commonfactors}-\ref{ass:contthreshvar}, \ref{ass:uniboundmom}, \ref{ass:p_thres_space}-\ref{ass:p_full_rank} and \ref{ass:NLP} hold, as $(N,T) \rightarrow \infty$ and $T/N \rightarrow 0$, under $H_{0}: \delta^0_i = 0$, for all $i$, then $supW \stackrel{d}{\rightarrow}  \mathop{sup}_{\gamma \in \Gamma}\mathcal{W}^c_{NT}(\gamma)$, where $  \mathcal{W}^c_{NT}(\gamma) = \bar{S}(\gamma)'\Bar{K}(\gamma,\gamma)^{-1}\bar{S}(\gamma)$, and $\Bar{S}(\gamma) = L'\Sigma(\gamma,\gamma)^{-1}S(\gamma)$ is a mean-zero Gaussian process with covariance kernel $\Bar{K}(\gamma^1,\gamma^2) = L'\Sigma(\gamma^1,\gamma^1)^{-1}K(\gamma^1,\gamma^2)\Sigma(\gamma^2,\gamma^2)^{-1}L$.
\end{theorem}

The asymptotic distributions described in both Theorems 9 and 10 depend on nuisance parameters. Therefore, as advocated by \citet{hansen1996inference}, \citet{chudik2017there}, and \citet{Greta2023}, we use the bootstrap method for inference. The steps for the bootstrap and Monte Carlo simulations evaluating its performance can be found in the supplementary online appendix. The $T/N\to 0$ assumption can be relaxed to $T/N\to \tau> 0$, by applying the bootstrap of \cite{DeVosStauskas2024}.

\section{Empirical Application}\label{section:application}
We apply the new theory to one of the key puzzles in international economics, namely the Feldstein-Horioka \citep{feldstein1980domestic}. In theory, perfect capital mobility should allow savings from one country to be invested in other countries where investment opportunities with higher returns are available. \citet{feldstein1980domestic} find however, that this is not the case and that domestic investments are highly correlated with domestic savings. Since then, the Feldstein-Horioka puzzle has become one of the six main puzzles of international macroeconomics \citep{obstfeld2000six}. \citet{feldstein1980domestic} used cross-sectional and time series data to estimate the relationship:
\begin{equation}
    \frac{I}{Y}=a+\beta\frac{S}{Y},
\end{equation}
where $Y$ is national income, $I$ is domestic investment, and $S$ is domestic savings. They estimate the savings retention rate $\beta$ close to $1$, rather than to $0$, which would apply in a world of perfect capital mobility. Furthermore, \citet{feldstein1980domestic} explored two additional issues: i) the existence of country-specific heterogeneity and ii) non-linearity with respect to trade openness. First, they empirically established that notable cross-country heterogeneity exists, as evidenced by the substantial variation in individual-country coefficient estimates. Second, they examined whether increased trade openness, which reduces economic frictions and hence facilitates capital mobility, would reduce the correlation between domestic savings and investment rates, as economic agents gain access to a broader array of global investment opportunities. Based on an interaction term between savings and trade openness, they found a minor and non-significant trade openness nonlinear effect.   

There is now a wide body of literature studying the existence of non-linearity and heterogeneity in the relationship between investment and savings. Recently, \cite{lusu23} examined the Feldstein-Horioka puzzle via a novel panel regression model with general forms of heterogeneity, where slope coefficients are allowed to vary over both individuals and time. However, this model considers only heterogeneity and not nonlinearity. Instead, \citet{haciouglu2021common} consider simultaneously heterogeneity and non-linearity based on trade openness, modeled via a smooth transition model based on the logistic function. Our analysis is closer to that of \citet{haciouglu2021common}, but differs in that we employ a discontinuous threshold model, and additionally, we uniquely allow for heterogeneous thresholds. 

The baseline model we consider is the fully heterogeneous one:
\begin{align} \label{emp_model}
    \textit{Investment}_{i,t} & =  \beta_{1i} \textit{Savings}_{i,t} + \beta_{2i} \textit{Trade Openness}_{i,t} \notag \\ 
    &   + \delta_i\textit{Savings}_{i,t}\mathbb{I}\{\textit{Trade Openness}_{i,t} \geq \gamma_i \}+ \alpha_i + \lambda_i' f_t + \varepsilon_{i,t},
\end{align}
where $\textit{Investment}_{i,t}$ is the investment share of real GDP per capita for country $i$ at year $t$, $\textit{Savings}_{i,t}$ is the percentage share of current savings to GDP per capita for country $i$ at year $t$, and $\textit{Trade Openness}_{i,t}$ denotes the trade openness for country $i$ at year $t$. The standalone trade-openness regressor is included to avoid potential omitted variable bias. 

To obtain potential efficiency benefits from pooling, we additionally consider the semi-homogeneous model with an alternative threshold variable, which has common support. We specify $\mathbb{I}\{p_i(\textit{Trade Openness}_{i,t}) \geq \gamma \}$, where $p_i(\gamma)$ is the percentile function of the distribution of $\textit{Trade Openness}_{i,t}$ across $t = 1,...,T$, for country $i$. Because the new threshold variable is a quantile, $\gamma$ is interpreted as the common quantile that separates the lower and upper regimes. To obtain the actual trade openness threshold for each country, we need to invert the quantile function. The common threshold is estimated at $\hat\gamma=68.5$ in the appendix; therefore, whenever trade openness in a specific country crosses the $68.5$ percentile of that particular country's distribution, that country crosses to the high regime. This interpretation is different from that of $\gamma_i$ in \eqref{emp_model} which are directly the levels of trade openness. 

In both models, $\alpha_i$ are the fixed effects, $\lambda_i' f_t $ are the interactive fixed effects, and $\varepsilon_{i,t}$, are the innovations. The individual effects $\alpha_i$ multiply $d_t=1$, for all $t$, which can be thought of as a ``known'' common factor, and as such can be treated differently from $f_t$. In this case, the matrix of cross-sectional averages becomes $\bar X^\star= [d, \bar X]$ where $d$ is a $T-$vector of ones. As mentioned in the discussion of the assumption \ref{ass:factorloadings}, common factors treated this way are not part of $f_t$ and do not count toward $m$, putting less strain on condition $m\leq K$. Additionally, we checked for variable non-stationarity using the CIPS test of \cite{Pesaran2007} which allows for a factor in the errors, and up to five lags of serial correlation in the errors. Investment and savings were found to be stationary at the 1\% level while trade openness was barely non-stationary, at the 10\% level, thus we have decided to maintain the model as is so that our results are comparable to the literature. Dynamics in the investment variable may also affect this regression. Given that lags of the dependent variable are not permitted as regressors, we estimate the static model without a lagged dependent variable, which is left as serial correlation in the errors. However, investment shocks may influence future values of the regressors through the unobserved common factors. Feedback in the form of $E(e_{i,t}x_{i,t+1})\neq 0$ is allowed because $e_{i,t}=\lambda_i'f_t+\varepsilon_{i,t}$ and $x_{i,t+1}=\Pi_i'f_{t+1}+\xi_{i,t+1}$ and hence feedback is present whenever $Cov(f_{t},f_{t+1})\neq 0$. The data is taken from Penn World Tables version 7.1 as in \citet{haciouglu2021common} and cover the period 1951-2000, resulting in a balanced panel with $N = 45$ and $T = 50$. 

The results of the MBIC criterion are $2.049$ for the fully heterogeneous model and $2.129$ for the semi-homogeneous model, selecting the fully heterogeneous model as the most appropriate. This is not surprising given the significant heterogeneity reported in \cite{feldstein1980domestic} and \cite{lusu23}. Therefore, in the following, we discuss the results only for the fully heterogeneous model and relegate the results for the semi-homogeneous model to the supplementary online appendix. 

\begin{table}[h]
\caption{Summary statistics for the estimated slope coefficients}\label{tableres}
\centering

\begin{tabular}{cccccccc}
\toprule\toprule
\multicolumn{8}{c}{$supW_i:$ Percentage of linearity test rejections at the 10\% level: 0.20} \\ \hline 
 Slope Coefficient & Mean &  St. Dev.   & $Q_1$   & $Q_2$   & $Q_3$     & Min     & Max    \\
\midrule
$\tilde \beta_{1i}$ & 0.722 & 0.426 & 0.502
 & 0.855  & 1.028  & -0.517  & 1.316 \\
$\tilde \beta_{2i}$           & 0.089 & 0.327
 & -0.053 & 0.096  & 0.239 & -0.749  & 0.876 \\
$\tilde \delta_i(^{*})$            & -0.128 & 0.335 & -0.357 & -0.208 & 0.037 & -0.504 & 0.617 \\
$\tilde \delta_i$ (all)            & -0.087 & 0.266 & -0.211 & -0.138 & 0.047 & -0.596 & 0.769 \\
$\hat \gamma_i$             & 51.684 & 40.687 & 26.752 & 38.105 & 63.868
 & 11.038 & 191.922 \\ 
\bottomrule
\end{tabular}
\begin{tablenotes}    
			\footnotesize           
			\item Notes: The table reports descriptive statistics for the estimates for the fully heterogeneous model \eqref{emp_model}. $\tilde \delta_i(^{*})$ refers to the $\tilde \delta_i$ coefficients from countries in which the test of linearity was rejected. 
		\end{tablenotes}  
\end{table}

The first line of Table \ref{tableres} presents results for the individual tests of nonlinearity. The vast majority, which is 80\% of the tests, do not reject the null hypothesis. The fifth line of the table contains descriptive statistics only for the 10\% statistically significant $\tilde{\delta}_{i}$. Compared to all $\tilde{\delta}_{i}$, which appear in the sixth line, the statistically significant ones are significantly larger in absolute value, in terms of both mean and median. In other words, when there is evidence for non-linearity, its effect is strong. For Cyprus, for example, $\tilde \beta_{1i}=0.957$, very close to one, indicating that investment depends largely on internal savings. However, $\tilde{\delta}_{i}=-0.34$ shows that once the country enters the high regime, the retention coefficient drops to $0.617$. 

When comparing the mean and median values of $\tilde\beta_{1i}$ and $\tilde\delta_{i}$, a form of analysis uniquely enabled by our model, we observe that the median offers even more evidence in favour of the Feldstein-Horioka puzzle. The median savings retention rate $\tilde\beta_{1i}$ is $0.855$, which is higher than the mean one. At the same time, the median $\tilde\delta$ is also larger in absolute value ($-0.208$) than the mean one, leading to a savings rate of $0.855-0.205=0.647$ in the high regime. Country-specific results can be found in the supplementary online appendix; for reference, the UK and the US have a retention coefficient of $0.855$ and $0.876$ respectively. France's $\tilde\beta_{1i}=1.085$, very close to $1.032$ estimated in \cite{feldstein1980domestic}. Luxembourg's retention coefficient is $0.289$, quite low, similar to $-0.298$ estimated in \cite{feldstein1980domestic}. In general, we observe that there is significant evidence in favour of the Feldstein-Horioka puzzle, but that the high trade openness regime leads to significant reductions to savings retention, for the countries for which there is evidence of nonlinearity. These results are broadly in line with the literature. The pooled estimates reported in \citet{feldstein1980domestic} and \citet{obstfeld2000six} are $0.89$ and $0.6$ respectively. \citet{haciouglu2021common} estimate $\tilde{\beta}_1$ at $0.69$ for the pooled estimator and $0.61$ for the mean-group, for OECD countries when the cross-sectional dependence is taken into account. \cite{lusu23} find a smaller estimate of $0.477$, without allowing for nonlinearity.   

The last line of the table contains summary statistics for the threshold variable $\gamma_i$. The average threshold is estimated at $51.684$, with the smallest threshold for India at $11.038$ and the largest threshold for Panama at $191.922$. The variance in these threshold values comes partly from the normalisation over GDP; in Panama imports and exports are large compared to their GDP, and the opposite holds for India.

Overall, the evidence in this paper can reconcile the results in papers such as \cite{feldstein1980domestic} that find no non-linearity and in \cite{haciouglu2021common}, that do find such evidence. The explanation we offer rests on country heterogeneity, where a few countries that experience threshold effects drive the results for the whole sample in pooled regression models. Countries that experience threshold effects include Cyprus, Ireland, Panama, Uruguay, and Japan. The first three countries are small economies that have been transformed by trade openness to become financial hubs. Therefore, external savings invested in these countries reduce the dependence of investment on domestic savings. Japan, on the other hand, has experienced prolonged low domestic returns on investment, encouraging Japanese investors to seek better returns in foreign markets. 
\section{Conclusion}
This paper proposes two models to accomodate heteterogeneity in panel threshold regression: one with heterogeneous slopes and thresholds, and another with heterogeneous slopes but homogeneous thresholds. Unobserved heterogeneity takes the form of IFE. We develop tests threshold effects, a criterion to choose between models, and an inferential theory for all parameter estimators. The new methods are validated by Monte Carlo simulations which can be found in the supplementary appendix. When applied to the Feldstein-Horioka puzzle the models show that cross-country heterogeneity is significant and that only a small subset of countries is responsible for previously reported trade openness nonlinearity.

There are still many interesting topics for future research. Interactive fixed effects represent a milestone in panel data analysis, and existing methods could be extended in this direction. Possible future research topics include panel threshold models with endogenous threshold variables as in \citet{seo2016dynamic}, multiple-regime threshold models, binary response models \citep{gao2023binary}, and quantile regression as in \citet{zhang2021single}.

\addcontentsline{toc}{section}{References}
\bibliographystyle{elsarticle-harv}
\bibliography{references}

\end{document}